\documentclass[
 reprint,
 superscriptaddress,
 amsmath,amssymb,
 aps,
]{revtex4-2}

\usepackage{graphicx}
\usepackage{dcolumn}
\usepackage{bm}
\usepackage[colorlinks=true, citecolor=Blue, linkcolor=Navy, urlcolor=RoyalBlue]{hyperref}

\usepackage{multirow} 
\newcolumntype{P}[1]{>{\centering\arraybackslash}p{#1}}
\newcolumntype{M}[1]{>{\centering\arraybackslash}m{#1}}

\usepackage{diagbox}
\usepackage[svgnames,table]{xcolor}

\begin{document}


\title{Expanding the Quantum-Limited Gravitational-Wave Detection Horizon}

\author{Liu~Tao}
\affiliation{Department of Physics \& Astronomy, University of California, Riverside, Riverside, CA 92521 USA}

\author{Mohak~Bhattacharya}
\affiliation{Department of Physics \& Astronomy, University of California, Riverside, Riverside, CA 92521 USA}

\author{Peter~Carney}
\affiliation{Department of Physics \& Astronomy, University of California, Riverside, Riverside, CA 92521 USA}

\author{Luis~Martin~Gutierrez}
\affiliation{Department of Physics \& Astronomy, University of California, Riverside, Riverside, CA 92521 USA}

\author{Luke~Johnson}
\affiliation{Department of Physics \& Astronomy, University of California, Riverside, Riverside, CA 92521 USA}

\author{Shane~Levin}
\affiliation{Department of Physics \& Astronomy, University of California, Riverside, Riverside, CA 92521 USA}

\author{Cynthia~Liang}
\affiliation{Department of Physics \& Astronomy, University of California, Riverside, Riverside, CA 92521 USA}

\author{Xuesi~Ma}
\affiliation{Department of Physics \& Astronomy, University of California, Riverside, Riverside, CA 92521 USA}

\author{Michael~Padilla}
\affiliation{Department of Physics \& Astronomy, University of California, Riverside, Riverside, CA 92521 USA}

\author{Tyler~Rosauer}
\affiliation{Department of Physics \& Astronomy, University of California, Riverside, Riverside, CA 92521 USA}

\author{Aiden~Wilkin}
\affiliation{Department of Physics \& Astronomy, University of California, Riverside, Riverside, CA 92521 USA}

\author{Jonathan~W.~Richardson}
\email{jonathan.richardson@ucr.edu}
\affiliation{Department of Physics \& Astronomy, University of California, Riverside, Riverside, CA 92521 USA}

\date{\today}

\begin{abstract}
We demonstrate the potential of new adaptive optical technology to expand the detection horizon of gravitational-wave observatories. Achieving greater quantum-noise-limited sensitivity to spacetime strain hinges on achieving higher circulating laser power, in excess of 1~MW, in conjunction with highly-squeezed quantum states of light. The new technology will enable significantly higher levels of laser power and squeezing in gravitational-wave detectors, by providing high-precision, low-noise correction of limiting sources of thermal distortions directly to the core interferometer optics. In simulated projections for LIGO~A+, assuming an input laser power of 125~W and an effective injected squeezing level of 9~dB entering the interferometer, an initial concept of this technology can reduce the noise floor of the detectors by up to 20\% from 200~Hz to 5~kHz, corresponding to an increment of 4~Mpc in the sky-averaged detection range for binary neutron star mergers. This work lays the foundation for one of the key technology improvements essential to fully utilize the scientific potential of the existing 4-km LIGO facilities, to observe black hole merger events past a redshift of~5, and opens a realistic pathway towards a next-generation 40-km gravitational-wave observatory in the United States, Cosmic~Explorer.
\end{abstract}

\maketitle


\section{Introduction\label{sec:intro}}

In the last decade, the Laser Interferometer Gravitational-Wave Observatory (LIGO) and the European Virgo observatory have established gravitational waves as a new observational probe of the universe, which carry complementary information to electromagnetic and particle observations. The current generation of detectors have now observed a wide variety of merger events involving black holes and neutron stars~\cite{GWTC-1, GWTC-2, GWTC-2.1, GWTC-3}. With further improvements in detector sensitivity, transformative new tests of strong-field gravity~\cite{Skenderis:2008, Cardoso:2017, Brustein:2018}, cosmology~\cite{Chen:2018, Farr:2019}, and dense nuclear matter~\cite{Tsang:2012} will become possible in the next decade. Realizing these improvements, however, requires significant new innovations in detector technology.

LIGO plans to incorporate such technological improvements in staged detector upgrades, beginning with LIGO~A+~\cite{Aplus} in 2025 and followed by LIGO~$\rm A^{\#}$~\cite{PostO5Report:2022} in 2029. The A+ upgrades are primarily aimed at reducing Brownian thermal noise on the mirror surfaces, through improved optical coatings. The $\rm A^{\#}$ upgrades will include larger 100-kg test masses, new improved suspensions, and a significantly higher laser power of 1.5~MW in the 4-km interferometer arms. LIGO~$\rm A^{\#}$ is envisioned to demonstrate most of the key technology for a next-generation 40-km gravitational-wave observatory in the United States, known as Cosmic Explorer~\cite{CEHorizonStudy}. With 320-kg test masses and a tenfold longer arm length, Cosmic Explorer will push the gravitational-wave detection horizon to near the edge of the observable universe.

\begin{figure*}
    \includegraphics[width=1\textwidth]{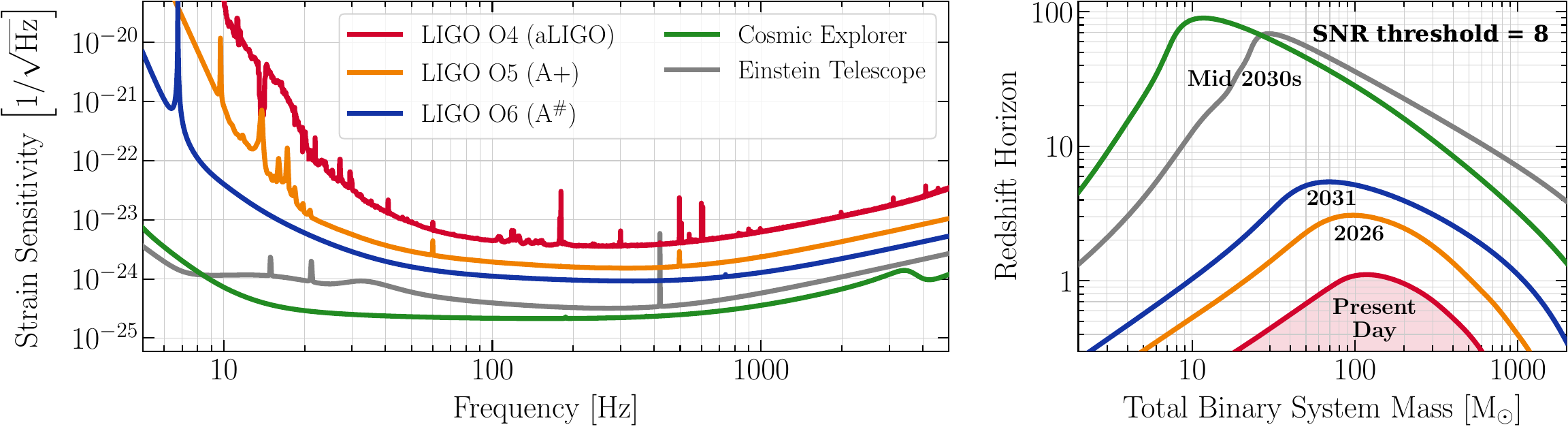}
    \caption{\label{fig:strain_targets}\textit{Left:} Strain sensitivity targets of each planned generation of U.S.-based gravitational-wave detectors. For reference, the next-generation European Einstein Telescope~\cite{EinsteinTelescope} is also shown. \textit{Right:} Corresponding cosmological redshift detection horizons for non-spinning, equal-mass binary black hole mergers, as a function of the total source-frame mass of the binary.}
\end{figure*}

Fig.~\ref{fig:strain_targets} compares the strain sensitivity targets of each successive detector generation (left panel), alongside the resulting improvement in astrophysical range as a function of source mass (right panel). With a factor of two sensitivity improvement over A+, the $\rm A^{\#}$ detectors will detect compact binary merger events past a redshift of~5, with annual detection rates of up to several thousand for binary black holes and several hundred for binary neutron stars. Cosmic~Explorer will achieve a further factor of five sensitivity improvement beyond $\rm A^{\#}$, extending the detection horizon to redshifts of up to~100, corresponding to a time when the Universe was only 0.1\% of its present age and resulting in the detection of hundreds of thousands to millions of compact binary events per year.

Achieving the sensitivities shown in Fig.~\ref{fig:strain_targets} hinges on reducing the quantum noise floor of gravitational-wave detectors. Quantum noise is the limiting source of instrumental noise across most of the frequency band accessible to ground-based detectors. It is limiting at all frequencies above 200~Hz. It arises not from the positional uncertainties of the interferometer's mirrors, or ``test masses,'' but from the quantization of the electromagnetic field used to interrogate their positions~\cite{Caves:1980}. Ground-state fluctuations of the electromagnetic vacuum field enter the interferometer and beat with the circulating laser field~\cite{Caves:1981}. At low frequencies (below 20~Hz), amplitude-quadrature fluctuations of the optical field are the most significant, which physically displace the test masses through radiation pressure. At higher frequencies, phase-quadrature fluctuations, manifesting as ``shot'' noise in the interferometer's readout, account for most of the quantum noise.

Quantum noise in gravitational-wave detectors can be reduced through two complementary means: (1)~higher circulating laser power in the interferometer and (2)~the injection of frequency-dependent ``squeezed'' quantum vacuum states~\cite{PhysRevD.104.062006, PhysRevX.13.041021}. To achieve their quantum noise targets, LIGO~$\rm A^{\#}$ and Cosmic Explorer will require 1.5~MW of laser power buildup in their arm cavities and 10~dB of observed squeezing~\cite{PostO5Report:2022, CEHorizonStudy}, roughly a fourfold increase beyond the highest levels achieved in Advanced LIGO. High laser power in fused-silica-based interferometers poses a major experimental challenge, due to the absorption of power by the mirror coatings and substrates. Absorption creates thermal gradients within the optics which induce optical aberrations~\cite{Brooks:2016}. They degrade both the optical power and squeezing, directly increasing the noise floor of the detector. Modeling has shown that sufficiently mitigating thermally-induced aberrations will require a \textit{qualitatively} new form of active wavefront correction on the test masses~\cite{AsharpTCSReqs:2022}, with actuation on smaller spatial scales (2-5~cm) than are accessible to LIGO's existing thermal compensation system (TCS)~\cite{Brooks:2016}.

In this letter, we present a novel adaptive optical concept that can deliver these essential new corrective capabilities. We show that it can enable significantly higher levels of laser power and squeezing in gravitational-wave detectors, opening a realistic pathway towards a next-generation observatory. In a companion paper~\cite{FROSTI_instrument}, we present experimental results from the testing of an actual prototype wavefront actuator on a 40-kg LIGO test mass, demonstrating the feasibility of this approach. This letter quantifies the projected \textit{impact} of this new technology on gravitational-wave astrophysics. We first overview our new adaptive optical technique in \S\ref{sec:frosti}. We then present the projected impact of an initial conception of this technology on LIGO~A+ in \S\ref{sec:procedure}-\ref{sec:impact}. Finally, we discuss the foundation this work lays for realizing future gravitational-wave detectors, beyond A+, in \S\ref{sec:outlook}.

\section{Next-Generation Adaptive Optics\label{sec:frosti}}

\begin{figure}[b]
    \centering
    \includegraphics[width=1\linewidth, trim={1mm 108mm 134mm 0mm}, clip]{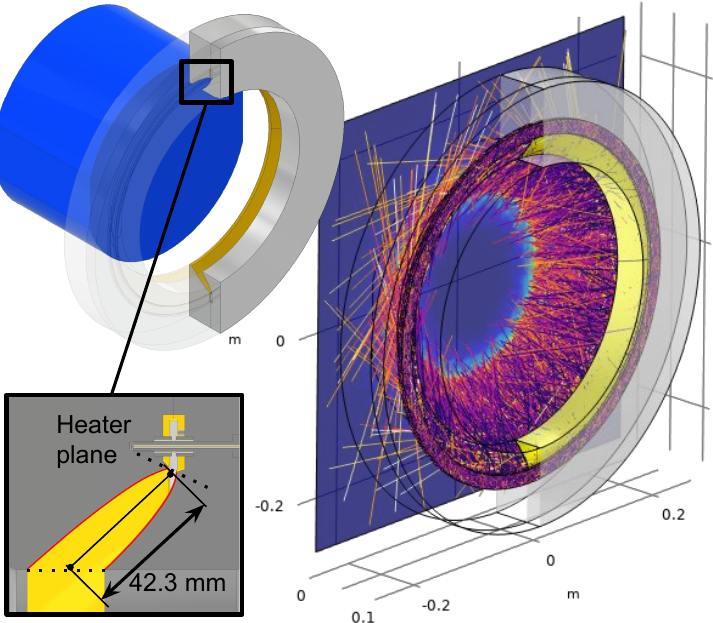}
    \caption{Ray-tracing simulation of the FROnt Surface Type Irradiator (FROSTI) designed to apply higher-order wavefront corrections to the LIGO test masses, by projecting an annular infrared radiation pattern onto their front surfaces. There the radiation is absorbed to provide corrective surface heating.}
    \label{fig:frosti_concept}
\end{figure}

\begin{figure*}[t]
    \centering
    \includegraphics[width=1\linewidth]{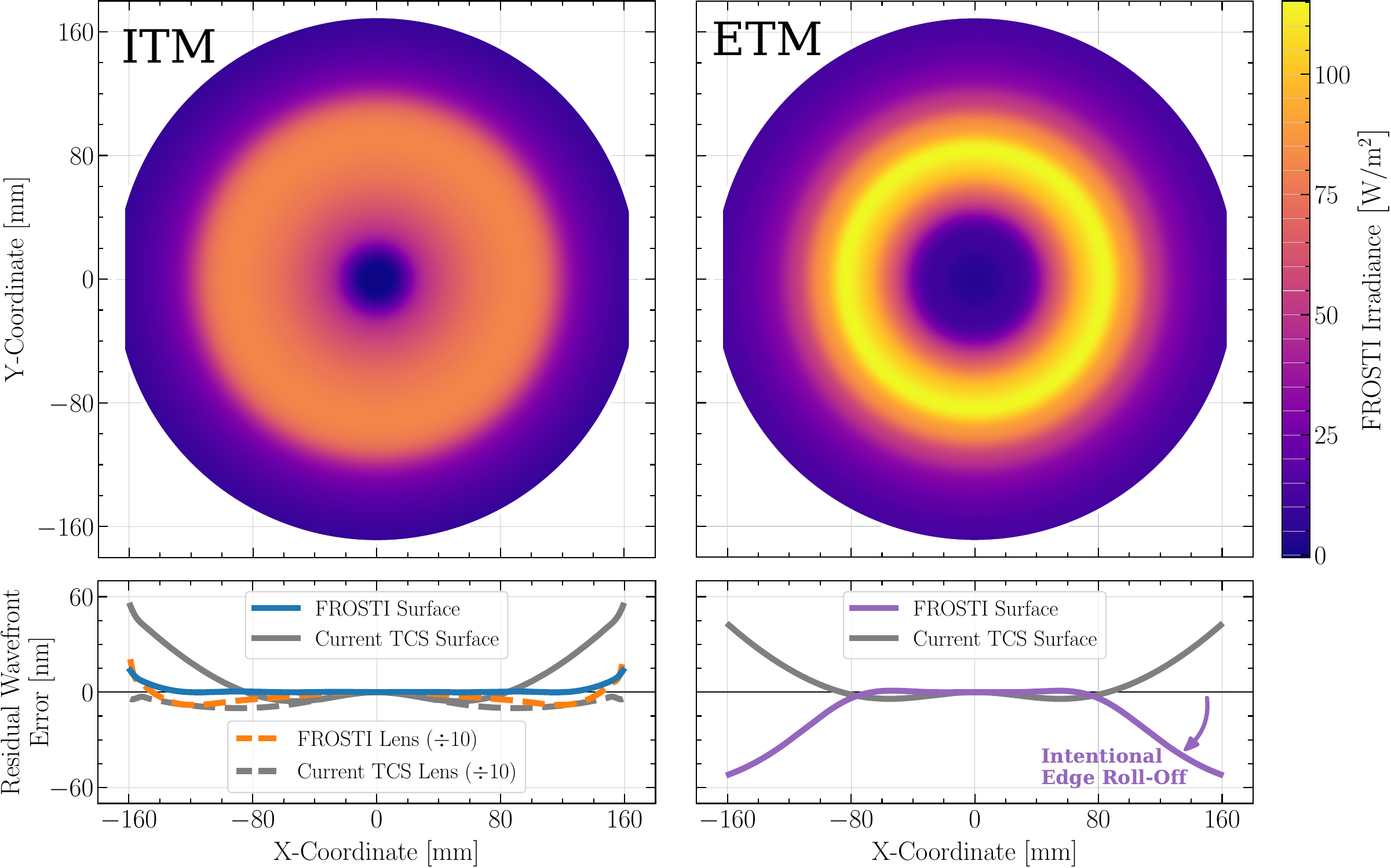}
    \caption{\textit{Top:} Optimal FROSTI irradiance profiles for the ITMs (left) and ETMs (right) in LIGO A+ obtained from numerical ray-tracing simulations with finite element analysis (after averaging along the azimuthal direction). \textit{Bottom:} Improvement in the residual wavefront errors due to thermoelastic surface deformation and thermorefractive substrate lensing, compared to the best achievable corrections with current thermal compensation system (TCS) capabilities. In the ETM case, additional surface roll-off near the edge is intentionally created to suppress the optical gain of problematic higher-order arm modes (HOM7)~\cite{Brooks:2021}.}
    \label{fig:TM_irradiances}
\end{figure*}

In future upgrades of LIGO and in Cosmic Explorer, the achievable laser power and squeezing will be limited by our ability to sense and correct thermal distortions of the optics. Thus, in practice, thermal distortions will impose the ultimate limit on the quantum-limited sensitivity of these detectors. Mitigating the residual, higher-order wavefront errors in the test masses, which arise from the mismatch between the distortion produced by the laser beam heating and the low-order (tilt and defocus) corrections that LIGO's TCS actuators can provide~\cite{AsharpTCSReqs:2022}, requires a new form of actuation on smaller 2--5~cm scales. Mitigating losses arising \textit{inside} the high-power arm cavities further requires that the higher-order wavefront correction be delivered in the form of a corrective heating profile applied directly to the front (reflective) surface of each test mass. A major challenge is that this correction must create negligible displacement noise. 

Intensity fluctuations of the incident heating beams will displace the test masses optomechanically and photothermally~\cite{Ballmer:2006, Willems:2011}, coupling actuator noise directly to the interferometer's readout signal. The most dominant of these noise couplings is flexure (bending) noise, thermoelastically-driven motion of the reflective surface relative to the test mass' center of gravity. For annular-like heating patterns, the requirements on their relative intensity noise (RIN) in A+ and A$^{\#}$ are $3 \times 10^{-8}/\sqrt{\text{Hz}}$ and $1 \times 10^{-8}/\sqrt{\text{Hz}}$, respectively, around 20~Hz~\cite{RINrequirement}. Blackbody radiation is currently the only technologically ready heating beam source with sufficient intensity stability to meet LIGO's stringent low-noise requirements.

The FROnt Surface Type Irradiator (FROSTI) is our concept for a new generation of wavefront actuators capable of meeting these challenging requirements. FROSTI actuators are designed to augment LIGO's existing TCS capabilities, by providing low-noise wavefront control on few-centimeter spatial scales. FROSTI is named for its function of restoring a test mass to its cold {\it optical} state (thermoelastically and thermorefractively) while illuminated by megawatt laser power. As illustrated in Fig.~\ref{fig:frosti_concept}, each actuator mounts in vacuum approximately 5~cm from the front surface of the test mass and just outside its 34-cm diameter. Its principle of operation is to project 3-14~$\mu$m blackbody radiation, produced by an internal ring heater and reshaped by nonimaging reflectors into a complex spatial pattern, onto the test mass surface. This radiation is fully absorbed within microns of the surface.

\begin{figure*}[t]
    \centering
    \includegraphics[width=1\linewidth]{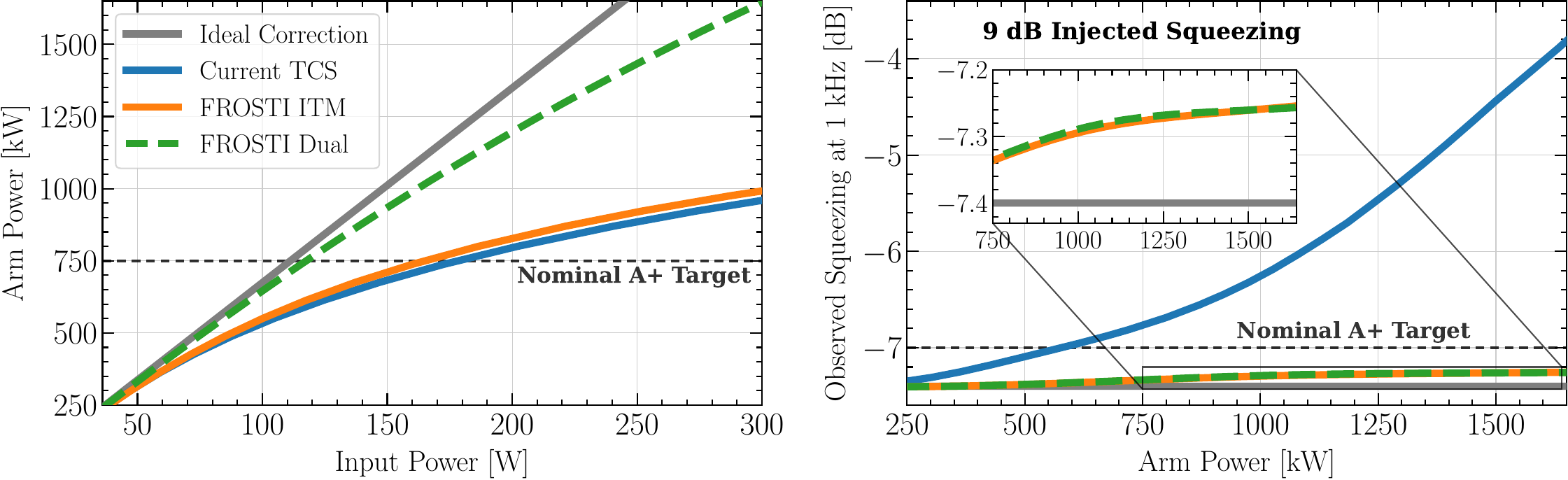}
    \caption{\textit{Left:} Achievable arm cavity power with increasing input power in the LIGO~A+ detectors, under different thermal compensation scenarios. \textit{Right:} Achievable squeezing (quantum noise reduction) with increasing arm power. In all cases, 9~dB of effective frequency-dependent squeezing enters the interferometer and the resulting \textit{observed} squeezing is shown at 1~kHz.}
    \label{fig:FROSTI_benefits_sqz} 
\end{figure*}

For each test mass, the radiation pattern is optimized using ray-tracing simulations incorporating a realistic design of the elliptical reflector surfaces, as described in our companion paper~\cite{FROSTI_instrument}, coupled to finite element analysis (FEA) models of the resulting thermoelastic and thermorefrective changes in the optic. Fig.~\ref{fig:TM_irradiances} shows the optimal irradiance profiles obtained for the A+ ITMs and ETMs from the FEA ray-tracing simulation. The ITM profile (top left panel), is optimized to simultaneously minimize the thermoelastic deformation of the reflective surface and the thermorefractive substrate lens, across the entire aperture of the optic. On the other hand, the ETM profile (top right panel) is optimized to minimize the thermoelastic deformation of its highly reflective (HR) surface near the center, where the majority of the fundamental laser mode's intensity is concentrated, while intentionally creating ``extra'' roll-off at the outer radii. This edge roll-off is designed to eliminate problematic higher-order mode co-resonances (HOM7) in the arm cavities, to additionally reduce the possible impacts of point absorbers in the future test mass coatings~\cite{Buikema:2020, Brooks:2021}. Since the ETM has a negligible power transmissivity (on the order of $10^{-6}$), only its surface profile, which determines the wavefront error imparted on the \textit{reflected} laser field, impacts the interferometer's performance. The final residual wavefront errors from the ITMs and ETMs are shown in Fig.~\ref{fig:TM_irradiances} (bottom panels), compared to the best possible corrections with current TCS capabilities.

\section{Simulation Procedure\label{sec:procedure}}

To study the potential impact of these new FROSTI actuators on LIGO, we utilize the frequency-domain interferometer simulation software \textsc{Finesse}~\cite{FinesseSoftware}. The \textsc{Finesse} simulations utilize Pound–Drever–Hall techniques~\cite{Drever:1983} to ``lock'' all of the relevant interferometer length degrees of freedom to the correct operating points, making use of the same radio-frequency sidebands and optical pick-offs as LIGO's length sensing and control system~\cite{Izumi:2017}. Frequency-dependent squeezing is injected from the anti-symmetric port, whose effective level is chosen so as to absorb all internal optical attenuation losses within the squeezer subsystem, injection optics, and the main interferometer optics. The only \textit{variable} source of squeezing loss is the thermally-induced mode-mismatch between the coupled cavities of the interferometer.

To self-consistently capture the limiting effects of thermal distortions as the interferometer is powered up, we couple the \textsc{Finesse} simulation to FEA models of the thermal distortions present, with increasing power levels, in each test mass. At each arm power level, we
\begin{enumerate}
    \vspace{-0.25\baselineskip}
    \item Rescale the mirror maps obtained from the FEA models of the thermal distortions in the test masses to reflect the interferometer's operating power.
    \vspace{-0.25\baselineskip}
    \item Re-optimize the power levels of the available TCS actuators at each test mass, to minimize the residual Gaussian-weighted RMS wavefront errors.
    \vspace{-0.25\baselineskip}
    \item Run the \textsc{Finesse} simulation with mirror maps corresponding to the updated self-heating and re-optimized TCS corrections for each test mass.
    \vspace{-0.25\baselineskip}
    \item Calculate the required \textit{input} power from the optical gains of the arm and power recycling cavities.
    \vspace{-0.25\baselineskip}
\end{enumerate}
The coating absorptivity of each test mass is assumed to be 0.5~ppm. Thus, each test mass absorbs 0.375~W at the nominal A+ arm power of 750~kW~\cite{Aplus}.

\begin{figure*}[t]
    \centering
    \includegraphics[width=1\linewidth]{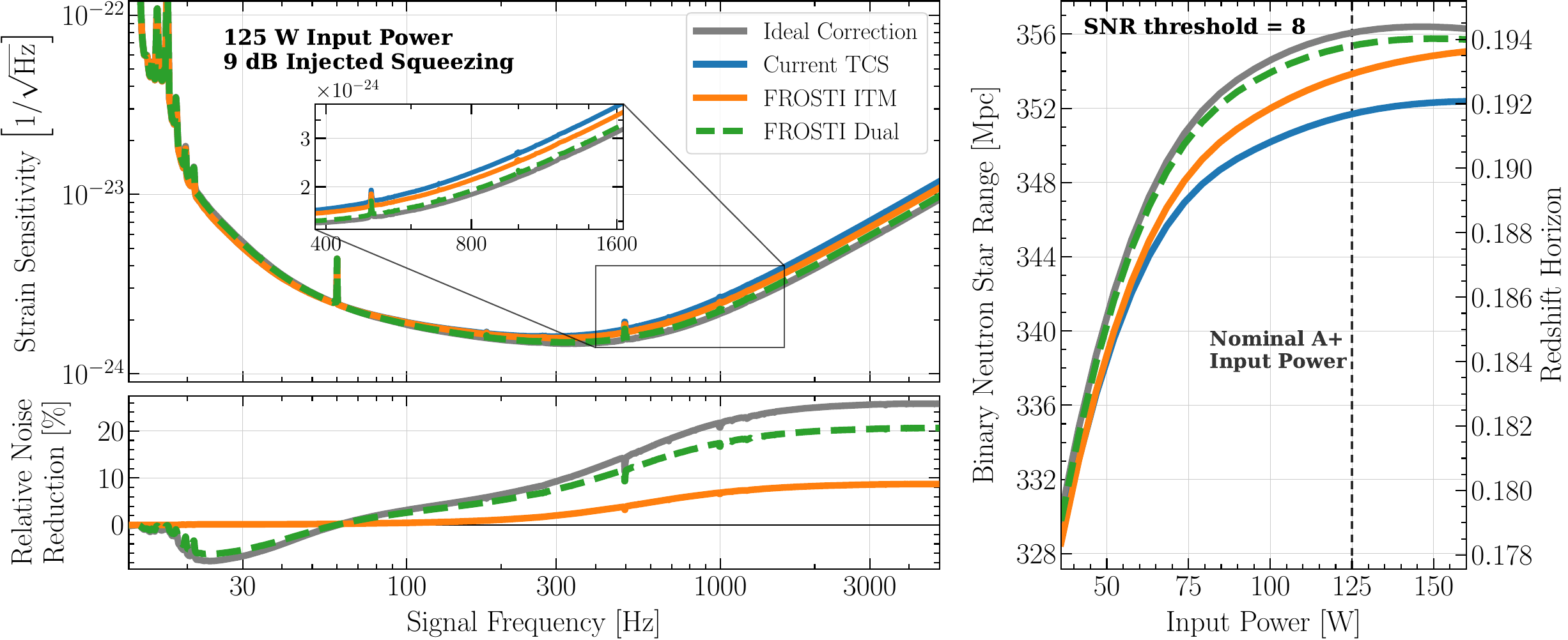}
    \caption{\textit{Left:} Projected strain sensitivity (top) and relative sensitivity improvement (bottom) of the LIGO~A+ detectors with 125~W of input laser power and 9~dB of effective injected squeezing, under different thermal compensation scenarios. The noise improvements in the bottom panel are calculated relative to the case of LIGO's current thermal compensation system (TCS). \textit{Right:} Astrophysical impact of the FROSTI-enabled quantum noise reduction in terms of binary neutron star detection range and the corresponding cosmological redshift horizon, as a function of increasing input power.}
    \label{fig:QN_sens_curves}
\end{figure*}

\section{Astrophysical Impact\label{sec:impact}}

Fig.~\ref{fig:FROSTI_benefits_sqz} shows the impact of the FROSTI actuators on the key performance metrics that determine LIGO's quantum-limited sensitivity. The left panel shows the projected arm cavity power, as a function of the input power, when the corrective heating profiles from Fig.~\ref{fig:TM_irradiances} are applied to only the ITMs (orange) or to both the ITMs and ETMs (green), compared to the best achievable correction with current TCS capabilities (blue). An ideal correction case where all thermal distortions are assumed to be perfectly corrected is also included as a reference (gray). As shown, we find that the dual-FROSTI correction enables LIGO to achieve the near-term A+ arm power target of 750~kW with a much lower requirement on the input laser power (120~W) than the current TCS, which requires 50\% more input power. It also enables LIGO to reach much higher arm power levels, up to the longer-term A$^{\#}$ arm power target of 1.5~MW, within the realistic limits of the available input laser power~\cite{PostO5Report:2022}. This is beyond the capability of the current TCS.

The right panel of Fig.~\ref{fig:FROSTI_benefits_sqz} shows the projected squeezing level (quantum noise reduction) at 1~kHz under each of these cases with increasing arm power. In each case, 9~dB of frequency-dependent squeezing is injected directly into the interferometer and the resulting squeezing \textit{observed} at the readout port is calculated. The effective level of injected squeezing accounts for all internal attenuation losses and phase noise from the squeezer subsystem and injection optics. The observed squeezing degradation is due to static attenuation losses from the main interferometer optics and readout optics (6\%), as well as dynamic thermally-induced mode-mismatch between the signal recycling cavity (SRC), arm cavities, and output mode cleaner (OMC). As shown, we find that both FROSTI cases (orange and green) maintain dramatically higher squeezing levels than the current TCS (blue), meeting the near-term A+ squeezing target of~7~dB while enabling higher-power operation. Thus, these actuators can overcome one of the greatest technical barriers to A$^{\#}$.

Finally, Fig.~\ref{fig:QN_sens_curves} shows how the enhanced thermal compensation capabilities provided by the FROSTI actuators lead to higher astrophysical sensitivity.  The left panels show the ultimate strain sensitivity of the LIGO~A+ detectors under each TCS actuation case (top) and the relative noise reduction compared to the current TCS case (bottom). In each case, we assume the nominal A+ input laser power of 125~W and 9~dB of effective injected squeezing, as well as the nominal A+ thermal noise and technical noise contributions. Across the 200~Hz to 5~kHz band limited by quantum shot noise, we find that the FROSTI actuators can enable up to a 20\% further reduction of the instrumental noise floor, relative to the current TCS. The relative sensitivity \textit{degradation} in the dual-FROSTI and ideal-correction cases at low frequencies arises from higher quantum radiation pressure noise due to their larger achieved arm powers. This is a real limitation for A+, but will be mitigated in A$^{\#}$ through its upgrade to larger 100-kg test masses. The right panel of Fig.~\ref{fig:QN_sens_curves} compares the sky-averaged detection range for binary neutron star mergers (left) and the corresponding redshift horizon (right), as a function of increasing input laser power. The optimal TCS actuator power levels for each case are listed in Table~\ref{tab-Aplusparameters}, at 125~W of input power.

\begin{table}[t]
    \centering
    \begin{tabular}{c|M{1.4cm}|M{0.9cm}|M{0.8cm}|M{0.9cm}|M{0.8cm}}
    \hline \hline 
    \multirow{2}{*}{\backslashbox{Optic}{Actuator}} & \multicolumn{3}{c|}{FROSTI ITM/Dual} & \multicolumn{2}{c}{Current TCS} \\
    \cline{2-6} & FROSTI & RH & CP & RH & CP \\
    \hline
    ITM & 4.2 & 16.2 & 0.0 &  4.6 & 3.3 \\
    \hline
    ETM & 4.7 & 13.3   & - & 3.5 & - \\
    \hline \hline
    \end{tabular}
    \caption{Optimal TCS actuator power levels, in units of watts, for the actuation cases shown in the left panel of Fig.~\ref{fig:QN_sens_curves}. Abbreviations: ring heater (RH); compensation plate (CP).}
    \label{tab-Aplusparameters}
    \vspace{-1\baselineskip}
\end{table}

\section{Future Directions\label{sec:outlook}}

Efforts are now underway to develop practical actuator designs for LIGO~A+ that deliver the ITM and ETM irradiance profiles presented in Fig.~\ref{fig:TM_irradiances}. These designs will be closely based on that of the demonstrated FROSTI prototype~\cite{FROSTI_instrument}. The impact of this work will also extend far beyond LIGO~A+, as it lays the foundation for a critical piece of next-generation gravitational-wave detector technology. Although the A+ detectors cannot fully benefit from the higher power levels enabled by FROSTI, due to increasing radiation pressure noise displacing its relatively light test masses, future extensions of this work will be integral to achieving the significantly higher power and squeezing targets of LIGO $\rm A^{\#}$ and Cosmic Explorer.

Preliminary modeling suggests that higher-precision wavefront corrections, capable of enabling even greater levels of quantum noise reduction, can be achieved with more complex heating patterns. Such irradiance profiles could be practically created by nesting \textit{multiple} heater rings inside one composite assembly, with each heater ring enclosed by a separate nonimaging reflector that projects the radiation onto a different radial zone of the test mass surface. This would allow the development of FROSTI actuators with multiple independent heating zones, enabling the creation of more complex irradiance profiles, and hence more precisely-targeted wavefront corrections, than can be produced with a single heater ring alone. Thus, the current work opens a key new research and development pathway towards realizing next-generation gravitational-wave observatories.

\vspace{-1\baselineskip}

\begin{acknowledgments}
The authors thank Aidan Brooks, Huy Tuong Cao, and Kevin Kuns for helpful discussions on the thermal FEA and quantum noise modeling. This material is based upon work supported by the National Science Foundation (NSF) under Award No. PHY-2110348. Additional support was provided by LIGO Laboratory under Advanced Detector Technology Research (ADTR) Initiative No. LIGO-M2200050. LIGO was constructed by the California Institute of Technology and Massachusetts Institute of Technology with funding from the NSF, and is operated by LIGO Laboratory under cooperative agreement PHY-1764464. This paper carries LIGO Document Number LIGO-P2400411.
\end{acknowledgments}

\section*{Data Availability}
Data underlying the results presented in this paper may be obtained from the authors upon reasonable request.

\bibliography{references}

\end{document}